\documentclass[article,prb,showpacs,preprintnumbers,amsmath,amssymb]{revtex4}
\usepackage{graphicx,epsf,dcolumn,subfigure}
\usepackage{placeins,longtable,fancyhdr}
\begin{document}



\begin{titlepage}

\def\baselinestretch{2.0}
\date{\today}

\title{Coulomb correlation effects in zinc monochalcogenides}

\author{S. Zh. Karazhanov}
\affiliation{Department of Chemistry, University of Oslo, PO Box
1033 Blindern, N-0315 Oslo, Norway}
\affiliation{Physical-Technical Institute, 2B Mavlyanov St.,
Tashkent, 700084, Uzbekistan}

\author{P. Ravindran}
\affiliation{Department of Chemistry, University of Oslo, PO Box
1033 Blindern, N-0315 Oslo, Norway} \email[Corresponding
author:]{ponniah.ravindran@kjemi.uio.no}

\author{U. Grossner}
\affiliation{Department of Physics, University of Oslo, PO Box
1048 Blindern, N-0316 Oslo, Norway}

\author{A. Kjekshus}
\author{H. Fjellv{\aa}g}
\affiliation{Department of Chemistry, University of Oslo, PO Box
1033 Blindern, N-0315 Oslo, Norway}

\author{B. G. Svensson}
\affiliation{Department of Physics, University of Oslo, PO Box
1048 Blindern, N-0316 Oslo, Norway}

\begin{abstract}

Electronic structure and band characteristics for zinc
monochalcogenides with zinc-blende- and wurtzite-type structures
are studied by first-principles density-functional-theory
calculations with different approximations. It is shown that the
local-density approximation underestimates the band gap and energy
splitting between the states at the top of the valence band,
misplaces the energy levels of the Zn-3$d$ states, and
overestimates the crystal-field-splitting energy. The
spin-orbit-coupling energy is found to be overestimated for both
variants of ZnO, underestimated for ZnS with wurtzite-type
structure, and more or less correct for ZnSe and ZnTe with
zinc-blende-type structure. The order of the states at the top of
the valence band is found to be anomalous for both variants of
ZnO, but is normal for the other zinc monochalcogenides
considered. It is shown that the Zn-3$d$ electrons and their
interference with the O-2$p$ electrons are responsible for the
anomalous order. The effective masses of the electrons at the
conduction-band minimum and of the holes at the valence-band
maximum have been calculated and show that the holes are much
heavier than the conduction-band electrons in agreement with
experimental findings. The calculations, moreover, indicate that
the effective masses of the holes are much more anisotropic than
the electrons. The typical errors in the calculated band gaps and
related parameters for ZnO originate from strong Coulomb
correlations, which are found to be highly significant for this
compound. The local-density-approximation with multiorbital
mean-field Hubbard potential approach is found to correct the
strong correlation of the Zn-3$d$ electrons, and thus to improve
the agreement between the experimentally established location of
the Zn-3$d$ levels and that derived from pure LDA calculations.

\end{abstract}



\maketitle

\end{titlepage}

\def\baselinestretch{2.0}

\normalsize
\section{\label{intro} Introduction}

Wide band-gap semiconductors are very important for applications
in optical devices such as visual displays, high-density optical
memories, transparent conductors, solid-state laser devices,
photodetectors, solar cells etc. The functional usefulness of such
devices of the zinc monochalcogenides depends on electronic
properties at the $\Gamma$~point at the valence-band (VB) maximum
and conduction-band (CB) minimum (recalling that these compounds
have direct band gaps). Therefore, first-principles calculations
for these compounds are of considerable importance.

Up to now, most \textit{ab initio} studies have been based on the
density-functional theory (DFT)\cite{HK64} in the local-density
approximation (LDA).\cite{KS65} For many materials, the theory
provides a good description of ground-state properties. However,
problems arise when the DFT--LDA approach is applied to materials
with strong Coulomb correlation
effects,\cite{ASKCS93,DBSHS98,BABH00} traceable back to the
mean-field character of the Kohn--Sham equations as well as to the
poor description of strong Coulomb correlation and exchange
interaction between electrons in the narrow $d$~band. One of the
problems is that the LDA error in calculation of the band gap
becomes larger than the common LDA error. Several attempts have
been made to include the correlation effects in the DFT--LDA
calculations. The LDA plus self-interaction correction
(LDA+$SIC$)\cite{SG90,LRLSM02,VKP95,VKP96,DPVASMACG04} eliminates
the spurious interaction of an electron with itself as occurring
in the conventional DFT--LDA. This approach has been widely used
to study compounds with completed
semicore-$d$~shells,\cite{LRLSM02,VKP95,VKP96,Q00,PZ03,DPVASMACG04}
and it is found to lower the Zn-3$d$ levels derived from the
simple LDA, thus giving better agreement with the measured X-ray
emission spectra (XES) and effective masses of
carriers.\cite{LRLSM02,VKP95,VKP96,PZ03} The calculated value of
the band gap ($E_g$) then falls within the established error
limits for the LDA.\cite{LRLSM02,VKP95,VKP96,AYA94,DPVASMACG04}

Another promising approach for correlated materials is the
so-called LDA plus multiorbital mean-field Hubbard potential
(LDA+$U$),\cite{ASKCS93,DBSHS98,BABH00} which includes the on-site
Coulomb interaction in the LDA Hamiltonian. After adding the
on-site Coulomb interaction to the L(S)DA Hamiltonian, the
potential becomes spin and orbital dependent. The LDA+$U$,
although being a mean-field approach, has the advantage of
describing both the chemical bonding and the electron--electron
interaction. The main intention of the LDA+$U$ approach is to
describe the electronic interactions of strongly correlated
states. Such a computational procedure is widely used to study
materials with ions that contain incomplete $d$ or $f$ shells,
e.g., transition-metal oxides, heavy fermion systems
etc.\cite{ASKCS93,DBSHS98,BABH00} Recently, such approach has been
applied to ZnO\cite{JW06,PM03,F04,KRGKFS052} with completed
semicore-Zn-3$d$~ valence shell. However, it is concluded in
Refs.~\onlinecite{PM03,F04} that LDA+$U$ calculations, in
principle, should not have improved the size of the $E_g$ because
the Zn-3$d$~bands are located well below the Fermi level
($E_{\rm{F}}$).

Despite the required large sized computations, different versions
of the GW approximation\cite{Hedin65,Hyb85} have also been used to
study zinc monochalcogenides.\cite{UHKS02,LICL02,OA00,RQNFS05}
(``G" stands for one-particle Green's function as derived from
many-body perturbation theory and ``W" for Coulomb screened
interactions.) This approximation can take into account both
non-locality and energy-dependent features of correlations in
many-body system and can correctly describe excited-state
properties of a system by including its ionization potential and
electron affinity. Band-structure studies using the the GW
correction show that $E_g$ is underestimated by 1.2~eV for
ZnO,\cite{UHKS02} by 0.53~eV for ZnS,\cite{LICL02} and by 0.55~eV
for ZnSe.\cite{LICL02} However, the GW calculations in
Ref.~\onlinecite{OA00} overestimated $E_g$ for ZnO by
$0.84~\rm{eV}$. Recent studies\cite{FH05} of Zn, Cd, and Hg
monochalcogenides by the GW approach has shown that the band-gap
underestimation is in the range 0.3--0.6~eV. Incorporation of the
plasmon--pole model for screening has lead to systematic errors.
Combination of exact exchange (EXX) DFT calculations in the
optimized-effective-potential approach with GW is
found\cite{RQNFS05} to give better agreement with the experimental
band gaps and the location of the Zn-3$d$ levels. Recently,
excellent agreement with experiment was achieved by all-electron
full-potential EXX calculations\cite{SDA05} for locations of the
$d$ bands for a number of semiconductors and insulators (Ge, GaAs,
CdS, Si, ZnS, C, BN, Ne, Ar, Kr, and Xe), although the band gap
was not as close to experimental data as found in other
pseudopotential EXX calculations.

Despite intense studies, many of the fundamental properties of
these materials are still poorly understood and further
experimental and theoretical studies are highly desirable. One
target is the so-called \textit{eigenvalue problem}. For example,
LDA underestimates \emph{E}$_g$ for
ZnO,\cite{OA00,UHKS02,PM03,F04,KRGKFS052} ZnS,\cite{LICL02} and
ZnSe,\cite{LICL02} by more (in fact by $> 50~\%$) than expected
for a typical LDA error. Also, the actual positions of the Zn-3$d$
levels,\cite{OA00,UHKS02,PM03,F04,KRGKFS052,LICL02} the band
dispersion, crystal-field splitting ($\Delta_{\rm{CF}}$), and
spin-orbit coupling splitting
($\Delta_{\rm{SO}}$)\cite{LRLSM02,KRGKFS052} are not reproduced
correctly. Neither the use of the generalized-gradient
approximation (GGA) nor the inclusion of SO coupling into the
calculations seems able to remedy the above
shortcomings.\cite{JSLH00,PM03,KRGKFS052}

The effective masses of the charge carriers are more indefinite
parameters for the zinc monochalcogenides. Owing to low crystal
quality only a few cyclotron resonance experiments have been
performed for ZnO,\cite{IOTTK01,OTAK96} ZnS,\cite{Madelung,XC93}
and ZnTe.\cite{CGGS79} The status for the present situation is
that effective masses from different \textit{ab initio} packages
and experiments scatter appreciably in publications on ZnO
(Refs.~\onlinecite{LRLSM02,XC93,IOTTK01,OTAK96,LYV96a,H73,OAIK02})
and ZnTe (Refs.~\onlinecite{Madelung,CGGS79,NF67}).

In this work zinc monochalcogenides (Zn$X,~X$=~O, S, Se, Te) in
the zinc-blende-(z-) and wurtzite-(w-)type structural arrangements
are studied by first-principles calculations within the LDA, GGA,
and LDA+$U$ approaches with and without SO coupling.

\section{Computational details}

The electronic band structure of the Zn$X$-z and -w phases is
studied using the VASP--PAW package\cite{vasp}, which calculates
the Kohn--Sham eigenvalues within the framework of DFT.\cite{HK64}
The calculations have been performed with the use of the
LDA,\cite{KS65} GGA,\cite{PBE96} and simplified rotationally
invariant LDA+$U$\cite{ASKCS93,DBSHS98} approaches. The exchange-
and correlation-energy per electron have been described by the
Perdew--Zunger parametrization\cite{PZ81} of the quantum Monte
Carlo procedure of Ceperley--Alder.\cite{CA80} The interaction
between electrons and atomic cores is described by means of
non-norm-conserving pseudopotentials implemented in the VASP
package.\cite{vasp} The pseudopotentials are generated in
accordance with the projector-augmented-wave (PAW)
method.\cite{PAW94,PAW99} For the PAW method, high-precision
energy cutoffs have been chosen for all the Zn$X$ phases
considered. The use of the PAW pseudopotentials addresses the
problem of the inadequate description of the wave functions in the
core region common to other pseudopotential
approaches.\cite{AFB01} The application allows us to construct
orthonormalized all-electron-like wave functions for the Zn-3$d$,
-4$s$ and anion-$s$ and -$p$ valence electrons of the zinc
monochalcogenides under consideration.

Self-consistent calculations were performed using a $10\times
10\times 10$~mesh frame according to Monkhorst--Pack scheme for
z-type structures and a similar density of $\bf{k}$ points in the
$\Gamma$-centered grids for w-type phases. The completly filled
semicore-Zn-3$d$ states have been considered as valence states.

For band-structure calculations we used the experimentally
determined crystal-structure parameters (Table~\ref{latparam}) for
all phases considered. The ideal positional parameter $u$ for $X$
in the w-type structures is calculated on the assumption of equal
nearest-neighbor bond lengths:\cite{LYV96a}
\begin{equation}\label{u}
    u = \frac{1}{3} \Bigl(\frac{a}{c}\Bigr)^2 + \frac{1}{4}
\end{equation}
The values of $u$ for the ideal case agree well with the
experimental values $u_{\rm{Expt}}$ (see Table~\ref{latparam}).
The unit-cell vectors of the z-type structures are
$\textbf{a}=(0,1/2,1/2)\textbf{a},
\textbf{b}=(1/2,0,1/2)\textbf{a},
\textbf{c}=(1/2,1/2,0)\textbf{a}$, $a$ is the cubic lattice
constant, and there are four Zn$X$ formula units per unit cell
specified by Zn at $(0,0,0)$ and $X$ at $(1/4,1/4,1/4)$. In the
w-type structure, the lattice vectors are
$\textbf{a}=(1/2,\sqrt{3}/2,0)\textbf{a},
\textbf{b}=(1/2,-\sqrt{3}/2,0)\textbf{a},
\textbf{c}=(0,0,c/a)\textbf{a}$, $c/a$ is the axial ratio, and
there are two Zn$X$ formula units per hexagonal unit cell, Zn at
$(0,0,0)$ and $(2/3,1/3,1/2)$ and $X$ at $(0,0,u)$ and
$(2/3,1/3,u+1/2)$.

The values of the $U$ and $J$ parameters were calculated within
the constrained DFT theory.\cite{PEE98} Furthermore, the position
of the Zn-3$d$ bands was calculated as a function of $U$ using the
LDA+$U$ method, and $U$ was derived semiempirically on forcing
match to the experimentally established\cite{RSS94} location of
the Zn-3$d$ bands. The thus obtained empirical values of $U$ were
used to explore further the electronic structure within the
LDA$+U$ procedure.

Values of $\Delta_{\rm CF}$, $\Delta_{\rm SO}$, and the average
band gap $E_0$ for ZnO (with anomalous order of the states at the
top of the VB) are calculated from the
expressions:\cite{H60,MRR95}
\begin{eqnarray}
E_0 & = & \frac{1}{3} \Bigl[ E_g(A) + E_g(B) + E_g(C)\Bigr] \\
{\Delta_{\rm CF} \atop \Delta_{\rm SO}} \Bigg\} & = & \frac{1}{2}
\Bigl[ \Delta_{\rm CB} - \Delta_{\rm BA} \pm \sqrt{2 \Delta_{\rm
CA}^2 -\Delta_{\rm BA}^2 - \Delta_{\rm CB}^2} \Bigr],
\label{CF-SO}
\end{eqnarray}
where $E_g(A), E_g(B),$ and $E_g(C)$ are energy gaps determined
from \textit{ab initio} calculations and $\Delta_{\rm CB} =
E_g(C)-E_g(B), \Delta_{\rm BA} = E_g(B)-E_g(A)$, and $\Delta_{\rm
CA} = E_g(C)-E_g(AB)$. To calculate these parameters for the other
Zn$X$ phases, and for ZnO with normal order of the states at the
top of the VB, $\Delta_{\rm CB}$ and $\Delta_{\rm CA}$ in
Eq.~\ref{CF-SO} have been exchanged.

For investigation of the order of the states at the top of the VB
for ZnO in z- and w-type structural arrangements, band-structure
calculations have been performed using the MindLab
package,\cite{MindLab04} which uses the full potential linear
miffin-tin orbital (FP-LMTO) method, and by the WIEN2K
code,\cite{WIEN2K01} which is based on a full-potential
linearized-augmented plane-wave method.

\section{\label{Results} Results and discussions}

\subsection{\label{VBstructure}The electronic structure at the top
of VB}

Important optical and transport properties for semiconductors are
determined by the carriers close to $\textbf{k}=\textbf{0}$ in the
vicinity of the $\Gamma$~point. The VB spectrum near the
$\Gamma$~point is different for z- and w-type materials. Without
SO coupling the top of the VB for phases with w-type structure is
split into a doublet $\Gamma_5$ and a singlet $\Gamma_1$ state by
the crystal field (Fig.~\ref{bandmixing}). The $\Gamma_5$ is a
$p_x, p_y$-like state, while $\Gamma_1$ is a $p_z$-like state.
Inclusion of SO coupling gives rise to three twofold degenerate
bands in the VB, which are denoted as $hh$ (heavy holes), $lh$
(light holes), and $sh$ (spin-split-off holes)
(Fig.~\ref{bandmixing}). These states correspond to $A$, $B$, and
$C$ exciton lines in photoluminescence experiments.\cite{MRR95}
The symmetry of two of these three bands are of $\Gamma_7$
character and one of $\Gamma_9$ character. The $\Gamma_7$ state
derived from $\Gamma_5$ will obtain a slight admixture of $p_z$
while $\Gamma_9$ stays unmixed $p_x$ and $p_y$ like. For ZnO,
these bands calculated within LDA, GGA, and LDA+$U$ for
$U<9.0$~eV, are in the order of decreasing energy $\Gamma_7,
\Gamma_9$, and $\Gamma_7$, which is referred to as
\textit{anomalous} order, resulting from a negative $\Delta_{\rm
SO}$.\cite{RCP68} For $U>9.0$~eV the lower $\Gamma_7$ state
interchanges with $\Gamma_9$, so the order becomes $\Gamma_7,
\Gamma_7$, and $\Gamma_9$. For the other Zn$X$-w phases the
sequence is $\Gamma_9, \Gamma_7, \Gamma_7$, named as normal
order,\cite{TH59} and the order was not changed by LDA+$U$.
Without SO coupling the VB spectrum near the $\Gamma$~point for
the Zn$X$-z phases originates from the sixfold degenerate
$\Gamma_{15}$ state. The SO interaction splits the $\Gamma_{15}$
level into fourfold degenerate $\Gamma_8$ ($hh$ and $lh$) and
doubly degenerate $\Gamma_7$ ($sh$) levels.


\subsection{Band structure}

Initial band-structure calculations have been performed for ZnO-w
using three different pseudopotentials for the oxygen atom
supplied with the VASP package: ordinary, soft, and high-accuracy
oxygen pseudopotentials. Band dispersion, band gaps $[E_g, E_g(A),
E_g(B),$ $E_g(C),$ and $E_0]$, $\Delta_{\rm CF}$, and $\Delta_{\rm
SO}$ corresponding to these oxygen pseudopotentials do not differ
significantly from each other and the subsequent calculations were
performed using the ordinary oxygen pseudopotential.

Band-parameter values calculated with or without taking SO
couplings into consideration are listed in Table~\ref{band-gap}.
Band gaps and the mean energy level of Zn-3$d$ electrons $E_d$
from LDA calculations are underestimated, while $\Delta_{\rm CF}$
is overestimated compared to the experimental data. The DFT-LDA
error is quite pronounced for ZnO compared to the other Zn$X$
phases and the discrepancy exceeds the usual error for LDA
calculations. The discrepancies in the calculated $\Delta_{\rm
CF}$ values for ZnO compared to experimental values are
unacceptably large. Except for ZnO, the calculated $\Delta_{\rm
CF}$ values within the different approaches do not differ much,
emphasizing that Coulomb correlation effects are more pronounced
for ZnO than ZnS, ZnSe, and ZnTe.

The calculations show (Table~\ref{band-gap}) that $\Delta_{\rm
SO}$ is much smaller than 1.0~eV for all phases except ZnSe-z
($\Delta_{\rm SO} = 0.40$~eV) and ZnTe-z ($\Delta_{\rm SO} =
0.97$~eV). The SO coupling energy calculated for ZnO-z and -w
within LDA and GGA, is negative, while it is positive for the
other Zn$X$ phases. The numerical value of $\Delta_{\rm SO}$
calculated within the three approaches considered came out close
to each other for all Zn$X$ phases. The numerical value of
$\Delta_{\rm SO}$ is severely underestimated for ZnO-w and ZnS-w
compared to experimental data. Our $\Delta_{\rm SO}$ values for
the other Zn$X$-z phases in Table~\ref{band-gap} are in good
agreement with theoretical calculations\cite{CW04} by the LAPW
method and the available experimental data.

\subsection{Density of states}

Analysis of the density of states (DOS) for the Zn$X$ phases
(Fig.~\ref{dos}), calculated within the LDA, shows that the
Zn-3$d$~states are inappropriately close to the CB, which
contradicts the findings from XPS, XES, and UPS
experiments.\cite{LPMKS74,VL71,RSS94} Furthermore, these states
and the top of the VB are hybridized. Distinct from the other
Zn$X$ phases considered, ZnO in both z- and w-type structure shows
artificially widened Zn-3$d$~states. These discrepancies indicate
strong Coulomb correlation between narrow Zn-3$d$ states which is
not accounted for correctly in the LDA calculations. As a
consequence, the interactions of the semicore-Zn-3$d$~states with
O-2$p$ in VB are artificially enlarged, the band dispersions are
falsified, the widths of the O-$p$ and Zn-3$d$ bands are altered,
and the latter are shifted inappropriately close to the CB. These
findings indicate that correlation effects of the Zn-3$d$ states
should be taken into account to obtain a more proper description
of the electronic structure for the Zn$X$ phases, especially for
the z- and w-type variants of ZnO.


The simplified rotationally invariant LDA$+U$
approach\cite{ASKCS93,DBSHS98} has been used to correct the strong
correlation of the Zn-3$d$ electrons. This approach uses $U$ and
$J$ to describe the strong Coulomb correlation, but since these
parameters do not explicitly take into account the final state, we
did not regard the regular LDA+$U$ approach as sufficiently
rigorous, and we therefore rather preferred empirically assigned
$U$ and $J$ values. For comparison, the values of $U$ and $J$ have
been calculated for some of the compounds within the constrained
DFT\cite{PEE98} (Table~\ref{U+J}), showing that the calculated
values to some extent agrees with those extracted semiempirically.

Using the semiempirical values for the parameters $U$ and $J$,
band-structure calculations have been performed within LDA+$U$.
Figure~\ref{fig:u} shows the dependence of the Zn-3$d$~mean level
($E_d$) and $E_g$ of the Zn$X$ phases on $U$. Analysis of the
illustrations show that the LDA+$U$-derived band gaps are more
reasonable than the pure LDA-derived band gaps (see also
Table~\ref{band-gap}). Moreover, the deviation of the $E_g$ values
obtained using LDA+$U$ from those obtained by experiments are much
smaller than those calculated using the pure LDA (Fig.~\ref{dos}
and Table~\ref{band-gap}).


The values of the peaks in the DOS (Fig.~\ref{dos}), corresponding
to the Zn-3$d$ states calculated by the LDA+$U$, are much larger
than those calculated using the pure LDA. This indicates that
according to the LDA+$U$ the semicore-Zn-3$d$ electrons become
more localized than according to the pure LDA. Distinct from the
other Zn$X$ phases considered, the width of the Zn-3$d$ bands for
ZnO calculated by LDA+$U$ become much narrower than that
calculated by LDA (Fig.~\ref{dos}). However, LDA+$U$ only slightly
changed the width of the Zn-3$d$ bands of the other Zn$X$ phases,
which leads one to conclude that the Coulomb correlation effects
for ZnO is more pronounced than for the other compounds
considered.

\subsection{Order of states at the top of VB}

The order of states at the top of VB in ZnO-w is a frequently
debated topic at present (see, e.g.,
Refs.~\onlinecite{RLJLCH99,LRLSM02,RLJC01}). The present project
has addressed this problem for the Zn$X$ phases by LDA, GGA, and
LDA+$U$ calculations within the VASP, MindLab, and WIEN2K packages
with and without including the SO couplings. The results obtained
by LDA and LDA+$U$ within VASP are presented in
Figs.~\ref{wFS}~and~\ref{zFS}. By inspection of the degeneracy of
the eigenvalues it is found that the normal order $\Gamma_5 >
\Gamma_1$ of the states at the top of VB is obtained by LDA
without SO coupling for all Zn$X$-w phases. The same order was
obtained by calculations within GGA. However, upon using the
LDA+$U$ approach with the semiempirical values for the parameter
$U$ (Table~\ref{U+J}) the order of the states at the top of VB for
the ZnO-w is changed, while there appears no changes for the other
Zn$X$ phases.



The variations in the order of the states at the top of the VB on $U$ are
systematically studied for the ZnO-z (Fig.~\ref{zZnO+FS}) and ZnO-w
(Fig.~\ref{wZnO+FS}) phases, with and without including SO coupling. It is found
that at $U \approx 9.0$~eV, the LDA+$U$ without SO coupling interchanges the
sequence of the VB states from $\rm{\Gamma}_5 > \rm{\Gamma}_1$ to $\rm{\Gamma}_1
> \rm{\Gamma}_5$.

Since one can treat the semicore-Zn-3$d$ electrons as core electrons and freeze
their interaction with VB in theoretical calculations, we studied the VB structure
of ZnO using the MindLab package\cite{MindLab04} including the semicore-Zn-3$d$
electrons in the core. On this assumption one obtains the order $\rm{\Gamma}_1
> \rm{\Gamma}_5$ at the top of the VB for ZnO-w. Hence, the order of the states
in this case can be traced back to the treatment of the Zn-3$d$ electrons. On
comparing the structures at the top of the VB calculated within the LDA and GGA
approaches it is found that only quantitatively small changes have occurred. Hence,
inhomogeneities in the electron gas do not affect the order of the states at the top
of VB, and only slightly change the band dispersion.



On the involvement of the SO coupling, the $\Gamma_5$ and
$\Gamma_1$ states of ZnO-w are split into two $\Gamma_7$ and one
$\Gamma_9$ states (see Fig.~\ref{bandmixing}). Orbital
decomposition analysis was performed to establish the origin and
order of these states using the band structure calculated
according to the WIEN2K package.\cite{WIEN2K01} The order of the
states was found to be $\Gamma_7>\Gamma_9>\Gamma_7$ (LDA, GGA, and
LDA+$U$ ($U<9.0$~eV) calculations), viz. ``anomalous"
order.\cite{RCP68,LRLSM02,LYV96a,T60,SMK65}

The above analysis shows that among the Zn$X$ phases considered,
ZnO are most sensitive to Coulomb correlation effects. The values
for $\Delta_{\rm{CF}}$ extracted from the $\Gamma_5$--$\Gamma_1$
splitting according to the LDA without SO coupling are positive
and decrease with increasing $U$ (Fig.~\ref{ZnO+CF+SO}).
Correspondingly, $\Delta_{\rm{SO}}$ obtained on including the SO
coupling came out negative and increased in size with increasing
$U$. The order of the states is in this case anomalous
($\Gamma_7>\Gamma_9>\Gamma_7$), thus supporting findings according
to the model of Thomas (see
Refs.~\onlinecite{RCP68,LRLSM02,LYV96a,SMK65,T60}).

At high values of $U$ ($> 9.0$~eV), $\Delta_{\rm{CF}}$ becomes
negative, which indicates inversion from $\Gamma_5>\Gamma_1$ to
$\Gamma_1 > \Gamma_5$ in the order of the states at the top of VB
without the SO coupling. Upon inclusion of the SO coupling the
order becomes $\Gamma_7>\Gamma_7>\Gamma_9$, which does not agree
with either of the two models for the order of states. At present
the safest conclusion is that the parameter $U$ takes a value
below $\sim$~9.0~eV and that the order of the states at the top of
VB consequently must be classified as anomalous.


The variation of the energy splitting of the $A, B,$ and $C$
states, expressed by $E_A-E_B, E_A-E_C,$ and $E_A-E_C,$ in the VB
for ZnO-z and -w as a function of $U$ is displayed in
Fig.~\ref{ZnO+CF+SO}. For values of $U$ below $9.0$~eV the energy
splitting decreases with increasing $U$. At higher values of $U,
E_A-E_B$ becomes negative and decreases, $E_B-E_C$ increases,
while $E_A-E_C$ stays more or less constant.


Distinct from ZnO-w, the other Zn$X$-w phases exhibit a normal order of the states
at the top VB ($\Gamma_5<\Gamma_1$ and $\Gamma_7<\Gamma_7<\Gamma_9$ without and with
the SO coupling, respectively). This order does not change upon variation of the
value of $U$ in the LDA+$U$ calculations.

Without introduction of the SO coupling, the top of the VB for the
Zn$X$-z phases is triple degenerate (see
Figs.~\ref{bandmixing}~and~\ref{zFS}). When the SO coupling is
included in the consideration, the VB maximum is split in fourfold
($\Gamma_8$) and twofold ($\Gamma_7$) states with the normal
$\Gamma_8>\Gamma_7$ order for $X$~=~S, Se, and Te. However, the
order is anomalous for ZnO-z ($\Gamma_7>\Gamma_8$) and in addition
$\Delta_{\rm{SO}}$ becomes negative. The dependence of
$\Delta_{\rm{SO}}$ on $U$ is shown in Fig.~\ref{zZnO+SO},
revealing that for $U>8.0$~eV $\Delta_{\rm{SO}}$ becomes positive,
and consequently that normal order is restored for the states at
the top of VB. So, one can ascribe the anomalous order of the
states at the top of VB in ZnO-z to Coulomb correlation effects
related to the Zn-3$d$ electrons.

For all cases considered, the GGA approximation did not influence
the order. Hence, inhomogeneities in the distribution of the
electron gas do not play a significant role for the order of the
states at the top of VB.


\section{Effective masses}

The CB states with short wave vectors ($\bf{k \approx 0}$) are
doubly degenerate with respect to spin and can be characterized by
one or two energy-independent effective masses for the z- and
w-type arrangements. The effective masses are calculated along the
directions $\Gamma \rightarrow A, \Gamma \rightarrow M$, and
$\Gamma \rightarrow K$ within the LDA, GGA, and LDA+$U$ approaches
with and without including the SO couplings
(Tables~\ref{wem}~and~\ref{zem}). According to the conventional
notations carrier effective masses for the Zn$X$-z phases are
distinguished by the indices $e$, $hh$, $lh$, and $sh$, $sh$
corresponding to $\rm{\Gamma}_1$, and $hh$ and $lh$ to
$\rm{\Gamma}_5$. The carrier masses for the Zn$X$-w phases are
distinguished by the indices $e$, $A$, $B$, and $C$.

The calculated $m_e$ for the Zn$X$-z phases are more isotropic
than those for the Zn$X$-w phases. The numerical values of $m_e$
for ZnO-w, ZnS-w, ZnSe-z, and ZnTe-z obtained by the LDA are
underestimated by about 50~$\%$ compared to experimental
findings,\cite{H73,Madelung,OTAK96,IOTTK01} while those for the
other Zn$X$ phases agree fairly well with experimental data. GGA
and LDA+$U$ calculations only slightly improved the LDA-derived
$m_e$ values for all Zn$X$ phases except ZnO, whereas the latter
showed much better agreement with LDA+$U$. This indicates once
again that correlation effects are more pronounced for ZnO than
for the other phases considered.

The electron effective mass is smaller along the direction $\Gamma
\rightarrow A (\|)$ than along $\Gamma \rightarrow M (\bot)$ and
$\Gamma \rightarrow K (\bot)$. This feature can be important in
film and superlattice constructions of these phases.\cite{XC93}
The heavy holes along all directions (see
Tables~\ref{wem}~and~\ref{zem}) and light holes along the $\Gamma
\rightarrow A (\|)$ direction are much heavier than other holes
and, in particular, CB electrons. For example, the carrier
transport in ZnO is dominated by electrons, while that by holes
can in practice be ruled out. This in turn explains the
experimentally established large disparity\cite{Madelung} between
electron and hole mobilities, and also may explain the large
optical non-linearity in ZnO.\cite{XC93} The effective masses of
the holes are more anisotropic than those of electrons, which can
be traced back to states at the top of VB associated with
O-$p$~orbitals, and this can give rise to anisotropy in parameters
like carrier mobility.\cite{S03}

On comparison of the $m_e$ values in
Tables~\ref{wem}~and~\ref{zem} one sees that the influence of SO
coupling on $m_e$ is very important for ZnSe-z and ZnTe-z, while
for the other phases its effect is small. The present values for
ZnO are in reasonable agreement with the experimental
values,\cite{IOTTK01,Madelung,H73,OTAK96} except for $m_A^{\|},
m_B^{\|}$, and $m_C^{\bot}$ (the latter discrepancies being not
understood) and in good agreement with those
obtained\cite{LRLSM02} by the FP-LMTO method.

\vspace{-0.5cm}
\section{Conclusions}
\vspace{-0.5cm}

Electronic structure and band characteristics for Zn$X$-z and -w
phases are studied by first-principles calculations within the
LDA, GGA, and LDA+$U$ approaches. It is found that LDA
underestimates the band gaps, the actual positions of the energy
levels of the Zn-3$d$ states,  and splitting energies between the
states at the top of the valence band, but overestimates the
crystal-field splitting energy. Spin-orbit coupling energy is
overestimated for ZnO-w, underestimated for ZnS-w, and comes out
more or less accurate for ZnS-z, ZnSe-z, and ZnTe-z.

The LDA+$U$ approach has been used to account properly for the
strong correlation of the Zn-3$d$ electrons. The value of the
Hubbard $U$ potential was varied to adjust the Zn-3$d$ band
derived from LDA toward lower energies and thus provide better
agreement with the experimentally established location of the
Zn-3$d$ levels from X-ray emission spectra. Using the $U$ values
obtained by this approach the calculated band gaps and band
parameters are improved according to the LDA+$U$ procedure
compared to the pure LDA approach.

The order of the states at the top of the valence band is
systematically examined for Zn$X$ phases. It is found that the
ZnO-z and -w phases exhibit negative SO splitting and anomalous
order of the states within LDA, GGA, and LDA+$U$ for $U<9.0$~eV,
and the model of Thomas\cite{RCP68,LYV96a,T60,LRLSM02} is
supported for these two Zn$X$ phases. It is found that in the
LDA+$U$ calculations the anomalous order is maintained up to $U
\approx\ 8.0$~eV for ZnO-z and $U \approx\ 9.0$~eV for ZnO-w. For
values of $U$ above these limits, the order is inverted. For
ZnO-w, $\Delta_{\rm{CF}}$ goes from positive to negative, whereas
$\Delta_{\rm{SO}}$ converts to a complex quantity, and becomes
thus meaningless.
Based on these analyses it is concluded that the Zn-3$d$ electrons
are responsible for the anomalous order of the states at the top
of the valence band in ZnO. In the other Zn$X$ phases considered,
the order is normal for all values of $U$ used in the
calculations. For the three approaches considered, our findings
confirm the model of Thomas\cite{T60} regarding the order of the
states in the valence band of ZnO.

Effective masses of electrons at the conduction band minimum and
of holes at the valence band maximum have been calculated along
the symmetry axis $\Gamma-M, \Gamma-A$, and $\Gamma-K$ for the
w-type phases and along $\Gamma-X, \Gamma-K$, and $\Gamma-L$ for
the z-type phases. Along the $c$ axis of the w-type phases the
light- and heavy-hole bands are degenerate, but the degeneracy is
broken when spin-orbit coupling is included. The heavy holes in
the valence band are found to be much heavier than the conduction
band electrons in agreement with experimental findings which show
higher electron mobility than hole mobility. The calculations,
moreover, reveal that effective masses of the holes are much more
anisotropic than those of the electrons. Conduction band electron
masses for ZnO-w, ZnS-w, ZnSe-z, and ZnTe-z calculated within LDA
are underestimated by about 50~$\%$ compared to experimental data,
while those for the other Zn$X$ phases are considered to agree
with experimental data.

The GGA approach did not remedy the DFT-LDA derived error in the
calculated energy gaps and band parameters. We found that
spin--orbit coupling is important for calculation of the
parameters for ZnSe-z and ZnTe-z, while it is not significant for
the other zinc monochalcogenides.

It should be noted that electronegativity difference (according to
the Pauling scale) 1.9 for ZnO is much larger than 0.9, 0.8, and
0.5 for ZnS, ZnSe, and ZnTe, respectively, which reflects that ZnO
is more ionic than the other Zn$X$ compounds. Consequently, our
calculated DOS for the topmost valence band is relatively narrow
for ZnO, which accordingly shows stronger correlation effects than
the other Zn$X$.

It is established that the unusually large errors in calculated
(according to DFT within LDA) band gaps and band parameters are
owing to strong Coulomb correlations, which are found to be most
significant in ZnO among the Zn$X$ phases considered. Also,
because of the increase in ionic radii of $X$ with increasing
atomic number the Zn--$X$ bond length systematically increases
from ZnO to ZnTe. As a result, the Zn-3$d$ band moves toward lower
energies (see Fig.~\ref{dos}) and behaves like core electrons. In
contrast, the relatively short Zn--O distance further confirms
that the interaction of the Zn-3$d$ electrons with the valence
band is stronger in ZnO than in the other Zn$X$ compounds.
Consistent with the above view point, the Zn-3$d$ band of ZnO-w
and ZnO-z is located closer to the topmost valence band, thus
increasing the influence of the Coulomb correlation effects to the
electronic structure compared to the other Zn$X$-w and Zn$X$-z
phases. The present conclusion is consistent with the results in
Ref.~\onlinecite{SDA05}, which report that locations of the $d$
bands of a number of semiconductors and insulators (Ge, GaAs, CdS,
Si, ZnS, C, BN, Ne, Ar, Kr, and Xe), determined from all-electron
full-potential exact-exchange-DFT calculations, are in excellent
agreement with experiment.

\vspace{-0.75cm}
\section*{Acknowledgments}
\vspace{-0.5cm} This work has received financial and
supercomputing support from FUNMAT. SZK has obtained partial
financial support from the Academy of Sciences of Uzbekistan
(Project N31-36). SZK also thanks R.~Vidya, P.~Vajestoon, and
A.~Klaveness (Department of Chemistry, University of Oslo, Oslo,
Norway) for discussions and assistance. The authors thank
Professor~M.A.~Korotin (Institute of Metal Physics, Ekaterinburg,
Russia) for help with the computations of the values of the
parameters $U$ and $J$ within the constrain DFT and Dr.~Karel
Knizek (Institute of Physics ASCR, Prague, Czech Republic) for
assistance with calculations according to the WIEN2K package.

\newpage

\newpage
\begingroup
\begin{table*}[h]
\caption{Experimentally determined unit-cell dimensions $a, c$,
volumes ($V$), and ideal ($u;$ calculated by Eq.~\ref{u}) and
experimental ($u_{\rm{Expt}}$) positional parameters for the $X$
atom of the w-type phases. For w-type structures $a=b$. For z-type
structures $a=b=c$ and all atoms are in fixed positions.}
\begin{ruledtabular}
\begin{tabular}{lddddd}
\multicolumn{1}{l}{Phase} & \multicolumn{1}{c}{$a$~(\AA)} & \multicolumn{1}{c}{$c$~(\AA)}  & \multicolumn{1}{c}{$V$~(\AA$^3$)} & \multicolumn{1}{c}{$u_{\rm{Expt}}$} & \multicolumn{1}{c}{$u$} \\
\hline
\ \\
ZnO-w\footnotemark[1]  & 3.250 & 5.207 & 47.62 & 0.3825 & 0.3799 \\
ZnS-w\footnotemark[2]  & 3.811 & 6.234 & 78.41 & 0.3750 & 0.3746 \\
ZnSe-w\footnotemark[3] & 3.996 & 6.626 & 91.63 & 0.3750 & 0.3712 \\
ZnTe-w\footnotemark[4] & 4.320 & 7.100 & 114.75& 0.3750 & 0.3734 \\
\ \\
ZnO-z\footnotemark[5]  & 4.620 &       & 98.61 &        &        \\
ZnS-z\footnotemark[6]  & 5.409 &       &158.25 &        &        \\
ZnSe-z\footnotemark[1] & 5.662 &       &181.51 &        &        \\
ZnTe-z\footnotemark[7] & 6.101 &       &227.09 &        &        \\
\end{tabular}
\end{ruledtabular} \label{latparam}
\footnotetext[1]{Experimental value from Ref.~\onlinecite{ICST01}.}
\footnotetext[2]{Experimental value from Refs.~\onlinecite{W86,XC93}.}
\footnotetext[3]{Experimental value from Refs.~\onlinecite{ICST01,Oleg94}.}
\footnotetext[4]{Experimental value from Refs.~\onlinecite{LRVRR03,TOM78}.}
\footnotetext[5]{Experimental value from Ref.~\onlinecite{BD54}.}
\footnotetext[6]{Experimental value from Refs.~\onlinecite{HM82,WLAB90}.}
\footnotetext[7]{Experimental value from Refs.~\onlinecite{AYA94,WLAB90}.}
\end{table*}
\endgroup

\newpage

\begingroup
\setlongtables \setcounter{LTchunksize}{10}
\setlength\LTcapwidth{6in}
\begin{longtable*}{llrrrrrrrrr}
\caption{Band gaps $[E_g, E_g(A), E_g(B), E_g(C),$ and $E_0]$,
crystal-field ($\Delta_{\rm CF}^0, \Delta_{\rm CF}$), and
spin-orbit ($\Delta_{\rm SO}$) splitting energies (all in eV) for
Zn$X$ phases with w- and z-type structures calculated within LDA,
GGA, and LDA+$U$ approaches. $E_g$ and $\Delta_{\rm CF}^0$ refer
to calculations without SO coupling, in all other calculations the
SO interactions are accounted for. Experimental values are quoted
when available.} \\
\hline \hline
\multicolumn{1}{l}{Phase} & \multicolumn{1}{l}{Method} & \multicolumn{1}{c}{$E_g$} & \multicolumn{1}{c}{$E_g(A)$} & \multicolumn{1}{c}{$E_g(B)$} & \multicolumn{1}{c}{$E_g(C)$} & \multicolumn{1}{c}{$E_0$} & \multicolumn{1}{c}{$E_d$} & \multicolumn{1}{c}{$\Delta^0_{\rm{CF}}$} & \multicolumn{1}{c}{$\Delta_{\rm{CF}}$} & \multicolumn{1}{c}{$\Delta_{\rm{SO}}$} \\
\hline
\endfirsthead

\caption[]{-- continued from previous page} \\
\hline \hline
\hline \multicolumn{1}{l}{Phase} & \multicolumn{1}{l}{Method} & \multicolumn{1}{c}{$E_g$} & \multicolumn{1}{c}{$E_g(A)$} & \multicolumn{1}{c}{$E_g(B)$} & \multicolumn{1}{c}{$E_g(C)$} & \multicolumn{1}{c}{$E_0$} & \multicolumn{1}{c}{$E_d$} & \multicolumn{1}{c}{$\Delta^0_{\rm{CF}}$} & \multicolumn{1}{c}{$\Delta_{\rm{CF}}$} & \multicolumn{1}{c}{$\Delta_{\rm{SO}}$} \\
\hline
\endhead

\hline \multicolumn{11}{c}{{Continued on next page}} \\ \hline
\endfoot

\hline \hline
\endlastfoot

ZnO-w   & \rm{LDA}     & 0.744 & 0.724 & 0.756 & 0.839 & 0.773 & $\sim$ 5.00  & 0.095 & 0.093 &-0.043 \\
        & \rm{GGA}     & 0.804 & 0.783 & 0.817 & 0.900 & 0.833 & $\sim$ 5.00  & 0.097 & 0.094 &-0.044 \\
        & \rm{LDA+$U$} & 1.988 & 2.008 & 2.053 & 2.053 & 2.038 & $\sim$ 10.00 &       &       &        \\
& Expt.\footnotemark[1]&&3.441 & 3.443 & 3.482 & 3.455 &              &       & 0.039 &-0.004 \\
ZnS-w   & \rm{LDA}     & 1.990 & 1.968 & 1.995 & 2.073 & 2.012 & $\sim$ 6.50  & 0.069 & 0.052 & 0.027 \\
        & \rm{GGA}     & 2.232 & 2.211 & 2.236 & 2.310 & 2.253 & $\sim$ 6.00  & 0.066 & 0.049 & 0.025 \\
        & \rm{LDA+$U$} & 2.283 & 2.260 & 2.286 & 2.366 & 2.304 & $\sim$ 8.20  & 0.059 & 0.055 & 0.026 \\
&Expt.\footnotemark[2] &&3.864 & 3.893 & 3.981 &       &              &       & 0.058 & 0.086 \\
&Expt.\footnotemark[2] &&3.872 & 3.900 &       &       &              &       & 0.006 & 0.092 \\
ZnSe-w  & \rm{LDA}     & 1.070 & 0.939 & 1.008 & 1.379 & 1.109 & $\sim$ 6.50  & 0.114 & 0.324 & 0.047 \\
        & \rm{GGA}     & 1.327 & 1.200 & 1.268 & 1.624 & 1.364 & $\sim$ 6.50  & 0.112 & 0.311 & 0.046 \\
        & \rm{LDA+$U$} & 1.404 & 1.271 & 1.334 & 1.721 & 1.442 & $\sim$ 9.30  & 0.101 & 0.347 & 0.041 \\
&Expt.\footnotemark[3] &&2.860 & 2.876 & 2.926 && $\sim$ 9.20 &       &       &        \\
ZnTe-w  & \rm{LDA}     & 1.052 & 0.760 & 0.820 & 1.691 & 1.091 & $\sim$ 7.50  & 0.086 & 0.838 & 0.033 \\
        & \rm{GGA}     & 1.258 & 0.974 & 1.032 & 1.875 & 1.294 & $\sim$ 7.20  & 0.084 & 0.812 & 0.032 \\
        & \rm{LDA+$U$} & 1.283 & 0.990 & 1.043 & 1.882 & 1.305 & $\sim$ 9.50  & 0.075 & 0.809 & 0.030 \\
& Expt.\footnotemark[4]&&2.260 &       &       &       &              &       &       &        \\
ZnO-z   & \rm{LDA}     & 0.573 & 0.555 &       & 0.588 & 0.577 & $\sim$ 4.60  &       &       &-0.033 \\
        & \rm{GGA}     & 0.641 & 0.615 &       & 0.649 & 0.638 & $\sim$ 4.60  &       &       &-0.034 \\
        & \rm{LDA+$U$} & 1.486 & 1.495 &       & 1.497 & 1.496 & $\sim$ 7.90  &       &       & 0.002 \\
        &\rm{Empirical}&       & 3.300 &       &       &       &              &       &       &       \\
ZnS-z   & \rm{LDA}     & 1.875 & 1.852 &       & 1.916 & 1.873 & $\sim$ 6.10  &       &       & 0.064 \\
        & \rm{GGA}     & 2.113 & 2.092 &       & 2.151 & 2.112 & $\sim$ 6.00  &       &       & 0.059 \\
        & \rm{LDA+$U$} & 2.332 & 2.310 &       & 2.389 & 2.336 & $\sim$ 9.00  &       &       & 0.079 \\
&Expt.\footnotemark[2] &&3.680 &       & 3.740 &       & $\geq$ 9.00  &       &       & 0.067 \\
&Expt.\footnotemark[2] &&3.780 &       & 3.850 &       &              &       &       &       \\
ZnSe-z  & \rm{LDA}     & 1.079 & 0.948 &       & 1.341 & 1.079 & $\sim$ 6.60  &       &       & 0.393 \\
        & \rm{GGA}     & 1.335 & 1.209 &       & 1.586 & 1.335 & $\sim$ 6.50  &       &       & 0.377 \\
        & \rm{LDA+$U$} & 1.421 & 1.291 &       & 1.700 & 1.427 & $\sim$ 9.05  &       &       & 0.409 \\
&Expt.\footnotemark[2] &&2.700 &       &       &       & $\sim$ 9.20  &       &       & 0.400 \\
&Expt.\footnotemark[2] &&2.820 &       &       &       &              &       &       & 0.400 \\
ZnTe-z  & \rm{LDA}     & 1.061 & 0.772 &       & 1.668 & 1.070 & $\sim$ 7.10  &       &       & 0.897 \\
        & \rm{GGA}     & 1.267 & 0.986 &       & 1.853 & 1.275 & $\sim$ 7.05  &       &       & 0.867 \\
        & \rm{LDA+$U$} & 1.329 & 1.046 &       & 1.956 & 1.349 & $\sim$ 9.90  &       &       & 0.911 \\
&Expt.\footnotemark[2] &&2.394 &       &       &       & $\sim$ 9.84  &       &       & 0.970 \\
&Expt.\footnotemark[2] &&      &       &       &       & $\sim$ 10.30 &       &       &       \\
\label{band-gap} \footnotetext[1]{Experimental value from
Ref.~\onlinecite{MRR95}.} \footnotetext[2]{Experimental value from
Ref.~\onlinecite{LB2}.} \footnotetext[3]{Experimental value from
Refs.~\onlinecite{LB2,Madelung}.} \footnotetext[4]{Experimental
value from Ref.~\onlinecite{ZYS02}.}
\end{longtable*}
\endgroup

\newpage

\begingroup
\begin{table*}[h]
\caption{Values of $U$ and $J$ calculated within the constrain
DFT\cite{PEE98} for Zn$X$-w phases and extracted within LDA+$U$ by
fitting the energy level of Zn-3$d$~electrons to band locations
from XPS, XES, and UPS experiments.\cite{LPMKS74,VL71,RSS94}
Calculations have not been performed for the z-Zn$X$ phases within
the constrained DFT.}
\begin{ruledtabular}
\begin{tabular}{lcdddddddd}
\multicolumn{1}{l}{Method} & \multicolumn{1}{c}{} & \multicolumn{1}{c}{ZnO-w} & \multicolumn{1}{c}{ZnS-w} & \multicolumn{1}{c}{ZnSe-w} & \multicolumn{1}{c}{ZnTe-w} & \multicolumn{1}{c}{ZnO-z} & \multicolumn{1}{c}{ZnS-z} & \multicolumn{1}{c}{ZnSe-z} & \multicolumn{1}{c}{ZnTe-z} \\
\hline
LDA+$U$       & $U$ & 13.00 & 6.00 & 8.00 & 7.00 & 8.00 & 9.00 & 8.00 & 8.00 \\
              & $J$ & 1.00  & 1.00 & 0.91 & 0.89 & 1.00 & 1.00 & 0.91 & 1.00 \\
Constrain DFT & $U$ & 11.10 & 9.74 & 9.33 & 7.00 &      &      &      &      \\
              & $J$ & 0.91  & 0.90 & 0.91 & 0.89 &      &      &      &      \\
\end{tabular}
\end{ruledtabular}  \label{U+J}
\vspace{8.0cm}
\end{table*}
\endgroup

\newpage

\begingroup
\setlongtables \setcounter{LTchunksize}{10}
\setlength\LTcapwidth{6in}
\begin{longtable*}{lldddddddd}
\caption{Effective masses of electrons and holes (in units of the
free-electron mass $m_0$) for Zn$X$-w calculated within LDA, GGA,
and LDA+$U$ approaches. The results are compared with the
calculated and experimental data from Ref.~\onlinecite{Madelung}
(directions not specified) and those calculated by FP~LMTO
(Ref.~\onlinecite{LRLSM02}), LCAO (Ref.~\onlinecite{XC93}) and
determined experimentally (Ref.~\onlinecite{H73}). Labelling of
the effective masses is not changed with changed order of the
states at the top of VB. } \\
\hline \hline
\multicolumn{1}{l}{Phase} & \multicolumn{1}{l}{Method} & \multicolumn{1}{c}{$m_e^{\|}$} & \multicolumn{1}{c}{$m_e^{\bot}$} & \multicolumn{1}{c}{$m_A^{\|}$} & \multicolumn{1}{c}{$m_A^{\bot}$} & \multicolumn{1}{c}{$m_B^{\|}$} & \multicolumn{1}{c}{$m_B^{\bot}$} & \multicolumn{1}{c}{$m_C^{\|}$} & \multicolumn{1}{c}{$m_C^{\bot}$} \\
\hline
\endfirsthead

\caption[]{-- continued from previous page} \\
\hline \hline
\multicolumn{1}{l}{Phase} & \multicolumn{1}{l}{Method} & \multicolumn{1}{c}{$m_e^{\|}$} & \multicolumn{1}{c}{$m_e^{\bot}$} & \multicolumn{1}{c}{$m_A^{\|}$} & \multicolumn{1}{c}{$m_A^{\bot}$} & \multicolumn{1}{c}{$m_B^{\|}$} & \multicolumn{1}{c}{$m_B^{\bot}$} & \multicolumn{1}{c}{$m_C^{\|}$} & \multicolumn{1}{c}{$m_C^{\bot}$} \\
\hline
\endhead

\hline \multicolumn{10}{c}{{Continued on next page}} \\
\hline
\endfoot

\hline \hline
\endlastfoot
\multicolumn{1}{l}{} & \multicolumn{1}{l}{} & \multicolumn{8}{c}{Without SO coupling} \\
ZnO-w & \rm{LDA}     & 0.139 & 0.132 & 2.943 & 2.567 & 2.943 & 0.150 & 0.157 & 3.476 \\
      & \rm{GGA}     & 0.147 & 0.140 & 3.233 & 2.864 & 3.233 & 0.162 & 0.161 & 2.272 \\
      & \rm{LDA}+$U$ & 0.234 & 0.221 &       & 4.770 &       &       & 3.750 & 0.266 \\
ZnS-w & \rm{LDA}     & 0.151 & 0.172 & 1.500 & 1.517 & 1.500 & 0.168 & 0.136 & 1.332 \\
      & \rm{GGA}     & 0.158 & 0.184 & 1.589 & 1.611 & 1.589 & 0.177 & 0.151 & 1.201 \\
      & \rm{LDA}+$U$ & 0.159 & 0.176 & 1.763 & 1.759 & 1.745 & 0.178 & 0.147 & 1.368 \\
ZnSe-w& \rm{LDA}     & 1.434 & 0.087 & 1.494 & 1.423 & 1.397 & 0.088 &       & 0.756 \\
      & \rm{GGA}     & 0.093 & 0.105 & 1.386 & 1.327 & 1.386 & 0.105 & 0.086 & 1.068 \\
      & \rm{LDA}+$U$ & 1.476 & 0.110 & 1.597 & 1.584 & 1.751 & 0.109 & 0.090 & 1.008 \\
ZnTe-w& \rm{LDA}     & 0.067 & 0.079 & 1.072 & 1.166 & 1.072 & 0.074 & 0.061 & 0.663 \\
      & \rm{GGA}     & 0.078 & 0.092 & 1.073 & 1.089 & 1.063 & 0.088 & 0.070 & 0.751 \\
      & \rm{LDA}+$U$ & 0.080 & 0.095 & 1.236 & 1.322 & 1.225 & 0.087 & 0.071 & 0.876 \\
\ \\
\multicolumn{1}{l}{} & \multicolumn{1}{l}{} & \multicolumn{8}{c}{With SO coupling} \\
ZnO-w & \rm{LDA}         & 0.137 & 0.130 & 2.447 & 2.063 & 2.979 & 0.227 & 0.169 & 0.288 \\
      & \rm{GGA}         & 0.144 & 0.143 & 2.266 & 0.351 & 3.227 & 0.300 & 0.165 & 0.537 \\
      & \rm{LDA}+$U$     & 0.189 & 0.209 & 0.207 &11.401 & 4.330 & 3.111 & 0.330 & 0.270 \\
&FP LMTO\footnotemark[1] & 0.230 & 0.210 & 2.740 & 0.540 & 3.030 & 0.550 & 0.270 & 1.120 \\
&Expt.\footnotemark[2]   & 0.24  &       & 0.590 & 0.590 & 0.590 & 0.590 & 0.310 & 0.550 \\
&LCAO\footnotemark[3]    & 0.280 & 0.320 & 1.980 & 4.310 &       &       &       &       \\
ZnS-w & \rm{LDA}         & 0.144 & 0.153 & 1.746 & 3.838 & 0.756 & 0.180 & 0.183 & 0.337 \\
      & \rm{GGA}         & 0.142 & 0.199 & 2.176 & 1.713 & 0.402 & 0.198 & 0.440 & 0.443 \\
      & \rm{LDA}+$U$     & 0.138 & 0.157 & 1.785 & 2.194 & 0.621 & 0.195 & 0.339 & 0.303 \\
&Expt.\footnotemark[4]   & 0.280 &       & 1.400 & 0.490 &       &       &       &       \\
&LCAO\footnotemark[3]    & 0.260 & 0.330 & 1.510 & 1.470 &       &       &       &       \\
ZnSe-w& \rm{LDA}         & 0.148 & 0.139 & 1.404 & 0.158 & 0.114 & 0.124 & 0.171 & 0.197 \\
      & \rm{GGA}         & 0.184 & 0.149 & 1.395 & 0.184 & 0.135 & 0.173 & 0.190 & 0.306 \\
      & \rm{LDA}+$U$     & 0.185 & 0.149 & 1.629 & 0.189 & 0.137 & 0.187 & 0.??? & 0.344 \\
ZnTe-w& \rm{LDA}         & 0.108 & 0.128 & 1.042 & 0.118 & 0.070 & 0.105 & 0.229 & 0.237 \\
      & \rm{GGA}         & 0.134 & 0.182 & 1.044 & 0.122 & 0.102 & 0.145 & 0.239 & 0.246 \\
      & \rm{LDA}+$U$     & 0.131 & 0.184 & 1.116 & 0.131 & 0.128 & 0.166 &       &       \\
&Expt.\footnotemark[4]   & 0.130 &       & 0.600 &       &       &       &       &       \\
\label{wem}
\footnotetext[1]{Theoretical value from Ref.~\onlinecite{LRLSM02}.}
\footnotetext[2]{Experimental value from Ref.~\onlinecite{H73}.}
\footnotetext[3]{Theoretical value from Ref.~\onlinecite{XC93}.}
\footnotetext[4]{Experimental value from Ref.~\onlinecite{Madelung}.}
\end{longtable*}
\endgroup

\newpage

\begingroup
\setlongtables \setcounter{LTchunksize}{10}
\setlength\LTcapwidth{7in}
\begin{longtable*}{lldddddddddd}
\caption{Effective masses of electrons and holes (in units of the
free-electron mass $m_0$) for Zn$X$-z. The results are compared to
calculated and experimentally determined data cited in
Ref.~\onlinecite{Madelung}. The labelling of the effective masses
for ZnO-z has not been changed upon the change in the order of the
states at the top of VB. }\\
\hline \hline
\multicolumn{1}{l}{Phase} & \multicolumn{1}{l}{Method} & \multicolumn{1}{c}{$m_e$} & \multicolumn{1}{c}{$m_{hh}^{100}$} & \multicolumn{1}{c}{$m_{hh}^{110}$} & \multicolumn{1}{c}{$m_{hh}^{111}$} & \multicolumn{1}{c}{$m_{lh}^{100}$} & \multicolumn{1}{c}{$m_{lh}^{110}$} & \multicolumn{1}{c}{$m_{lh}^{111}$} & \multicolumn{1}{c}{$m_{SO}^{100}$} & \multicolumn{1}{c}{$m_{SO}^{110}$} & \multicolumn{1}{c}{$m_{SO}^{111}$} \\
\hline
\endfirsthead

\caption[]{-- continued from previous page} \\
\hline \hline
\multicolumn{1}{l}{Phase} & \multicolumn{1}{l}{Method} & \multicolumn{1}{c}{$m_e^{\|}$} & \multicolumn{1}{c}{$m_e^{\bot}$} & \multicolumn{1}{c}{$m_A^{\|}$} & \multicolumn{1}{c}{$m_A^{\bot}$} & \multicolumn{1}{c}{$m_B^{\|}$} & \multicolumn{1}{c}{$m_B^{\bot}$} & \multicolumn{1}{c}{$m_C^{\|}$} & \multicolumn{1}{c}{$m_C^{\bot}$} \\
\hline
\endhead

\hline \multicolumn{12}{c}{{Continued on next page}} \\
\hline
\endfoot

\hline \hline
\endlastfoot

\hline \multicolumn{1}{l}{} & \multicolumn{1}{l}{} &
\multicolumn{10}{c}{Without SO coupling}
\ \\
ZnO-z & \rm{LDA}            & 0.110  & 1.400  & 5.345  & 2.738  & 1.400  & 1.436  & 2.738 & 0.120  & 0.114 & 0.112 \\
      & \rm{GGA}            & 0.120  & 1.480  & 5.800  & 3.162  & 1.480  & 1.540  & 3.162 & 0.136  & 0.130 & 0.125 \\
      & \rm{LDA}+$U$        & 0.193  & 1.780  & 8.041  & 3.820  & 1.780  & 1.727  & 3.820 & 0.224  & 0.202 & 0.198 \\
ZnS-z & \rm{LDA}            & 0.155  & 0.662  & 3.405  & 1.467  & 0.662  & 0.683  & 1.467 & 0.161  & 0.134 & 0.129 \\
      & \rm{GGA}            & 0.172  & 0.710  & 3.800  & 1.500  & 0.710  & 0.710  & 1.500 & 0.188  & 0.145 & 0.155 \\
      & \rm{LDA}+$U$        & 0.177  & 1.674  & 4.318  &        & 1.674  & 0.882  &       & 0.214  & 0.164 & 0.192 \\
ZnSe-z& \rm{LDA}            & 0.084  & 0.606  & 3.520  & 1.383  & 0.606  & 0.585  & 1.383 & 0.085  & 0.076 & 0.076 \\
      & \rm{GGA}            & 0.100  & 0.626  & 3.430  & 1.320  & 0.626  & 0.600  & 1.320 & 0.106  & 0.090 & 0.090 \\
      & \rm{LDA}+$U$        & 0.097  & 0.667  &        & 1.605  & 0.677  & 0.677  & 1.605 & 0.105  & 0.090 & 0.090 \\
ZnTe-z& \rm{LDA}            & 0.073  & 0.445  & 2.764  & 1.003  & 0.451  & 0.450  & 1.042 & 0.073  & 0.065 & 0.062 \\
      & \rm{GGA}            & 0.085  & 0.440  & 2.701  & 1.046  & 0.443  & 0.452  & 1.044 & 0.086  & 0.075 & 0.072 \\
      & \rm{LDA}+$U$        & 0.090  & 0.519  & 3.812  & 1.202  & 0.519  & 0.516  & 1.202 & 0.089  & 0.075 & 0.075 \\
\ \\
\multicolumn{1}{l}{} & \multicolumn{1}{l}{} &
\multicolumn{10}{c}{With SO coupling}
\ \\
ZnO-z & \rm{LDA}       & 0.110  & 0.390  & 0.571  & 0.385  & 1.520 & 1.100  & 1.330  & 0.174  & 0.164 & 0.169 \\
      & \rm{GGA}       & 0.120  & 0.409  & 0.579  & 0.492  & 1.505 & 1.252  & 1.281  & 0.188  & 0.186 & 0.181 \\
      & \rm{LDA}+$U$   & 0.193  & 1.782  & 2.920  & 1.972  & 0.968 & 1.392  & 1.669  & 0.250  & 0.240 & 0.230 \\
ZnS-z & \rm{LDA}       & 0.150  & 0.775  & 1.766  & 2.755  & 0.224 & 0.188  & 0.188  & 0.385  & 0.355 & 0.365 \\
      & \rm{GGA}       & 0.172  & 0.783  & 1.251  & 3.143  & 0.233 & 0.216  & 0.202  & 0.378  & 0.373 & 0.383 \\
      & \rm{LDA}+$U$   & 0.176  & 1.023  & 1.227  & 1.687  & 0.268 & 0.252  & 0.218  & 0.512  & 0.445 & 0.447 \\
&Expt.\footnotemark[1] & 0.184  &        &        & 1.760  &       & 0.230  &        &        &       &       \\
&Expt.\footnotemark[1] & 0.340  &        &        &        &       &        &        &        &       &       \\
ZnSe-z& \rm{LDA}       & 0.077  & 0.564  & 1.310  & 1.924  & 0.104 & 0.100  & 0.094  & 0.250  & 0.246 & 0.254 \\
      & \rm{GGA}       & 0.098  & 0.568  & 0.922  & 1.901  & 0.126 & 0.122  & 0.111  & 0.271  & 0.273 & 0.267 \\
      & \rm{LDA}+$U$   & 0.100  & 0.636  & 1.670  & 1.920  & 0.129 & 0.120  & 0.117  & 0.287  & 0.297 & 0.309 \\
&Expt.\footnotemark[1] & 0.130  & 0.570  & 0.750  &        &       &        &        &        &       &       \\
&Expt.\footnotemark[1] & 0.170  &        &        &        &       &        &        &        &       &       \\
ZnTe-z& \rm{LDA}       & 0.064  & 0.381  & 0.822  & 1.119  & 0.071 & 0.067  & 0.066  & 0.254  & 0.253 & 0.256 \\
      & \rm{GGA}       & 0.078  & 0.418  & 0.638  & 1.194  & 0.093 & 0.086  & 0.081  & 0.261  & 0.255 & 0.274 \\
      & \rm{LDA}+$U$   & 0.081  & 0.483  & 0.929  & 1.318  & 0.096 & 0.088  & 0.085  & 0.288  & 0.292 & 0.290 \\
&Expt.\footnotemark[1] & 0.130  &        & 0.600  &        &       &        &        &        &       &       \\
\label{zem} \footnotetext[1]{Experimental value from Ref.~\onlinecite{Madelung}.}
\end{longtable*}
\endgroup

\newpage

{\textbf{Figure captions}

\begin{figure}[h]
\vspace{-5.0cm} \centering
\includegraphics[scale=0.75]{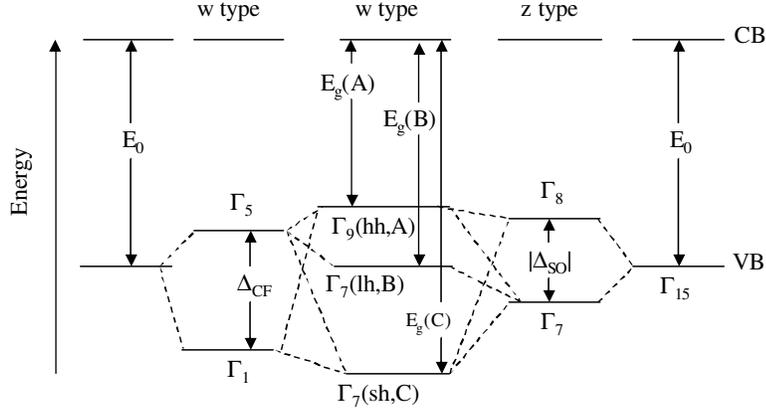}
\setlength{ \abovecaptionskip}{-12.0cm} \caption{Schematic
representation of the band mixing in Zn$X$ phases with z- and
w-type structure. In w-type structures the levels $\Gamma_9$,
$\Gamma_7$ (upper), and $\Gamma_7$ (lower) are formed due to the
combined influence (in the middle) of $\Delta_{\rm CF}$ (on the
left) and $\Delta_{\rm SO}$ (on the right). In z-type phases the
levels $\Gamma_8$ and $\Gamma_7$ are separated due to the SO
interaction.} \label{bandmixing}
\end{figure}

\begin{figure}[h]
\centering
\includegraphics[scale=1.0]{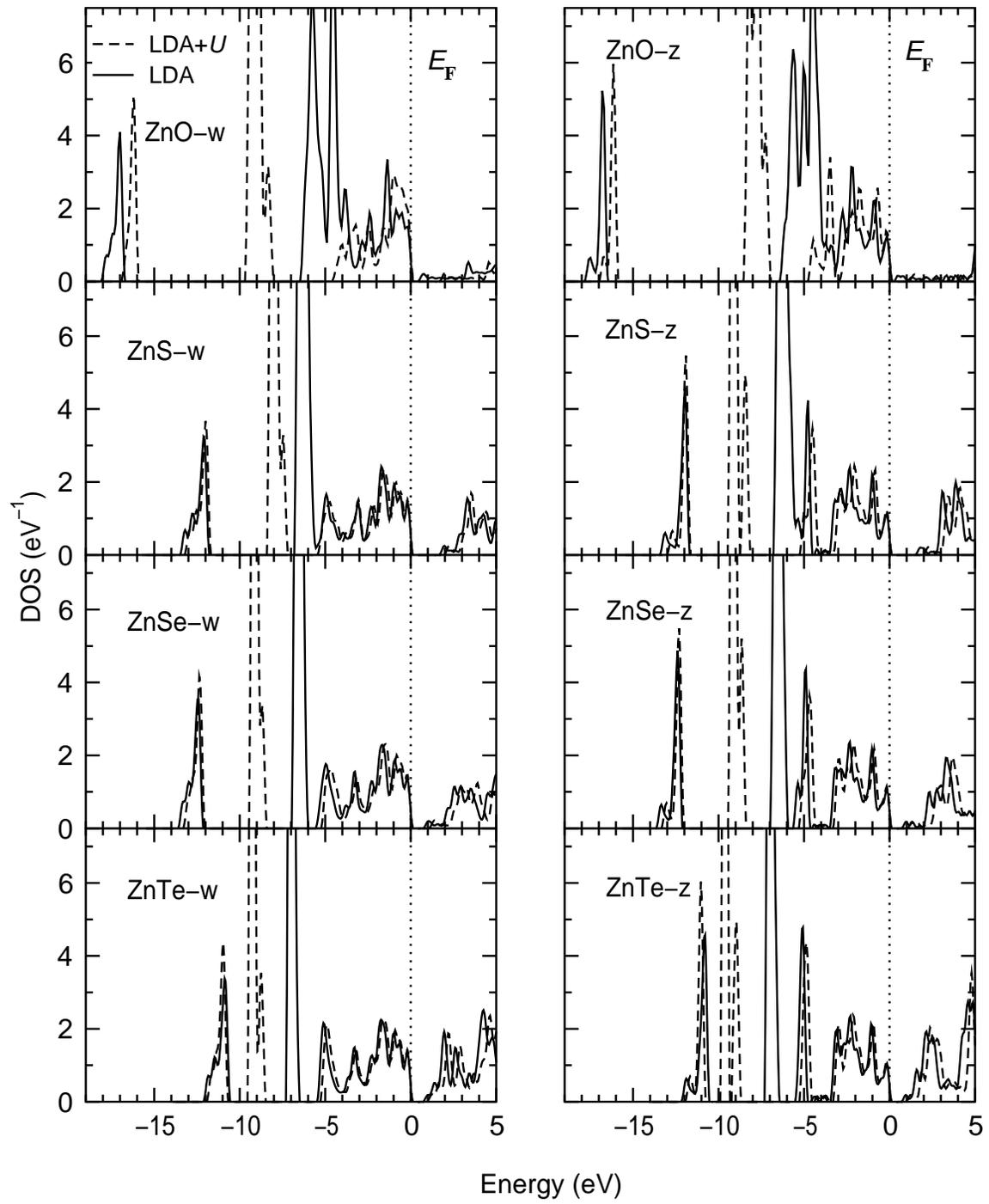}
\caption{Total density of states for Zn$X$-z and -w phases
calculated from the LDA (solid line) and LDA+$U$ (broken line)
approaches. $E_{\rm{F}}$ is marked by the dotted line.}
\label{dos}
\end{figure}

\begin{figure}[h]
\centering
\vspace{-1.0cm}
\includegraphics[scale=1.5]{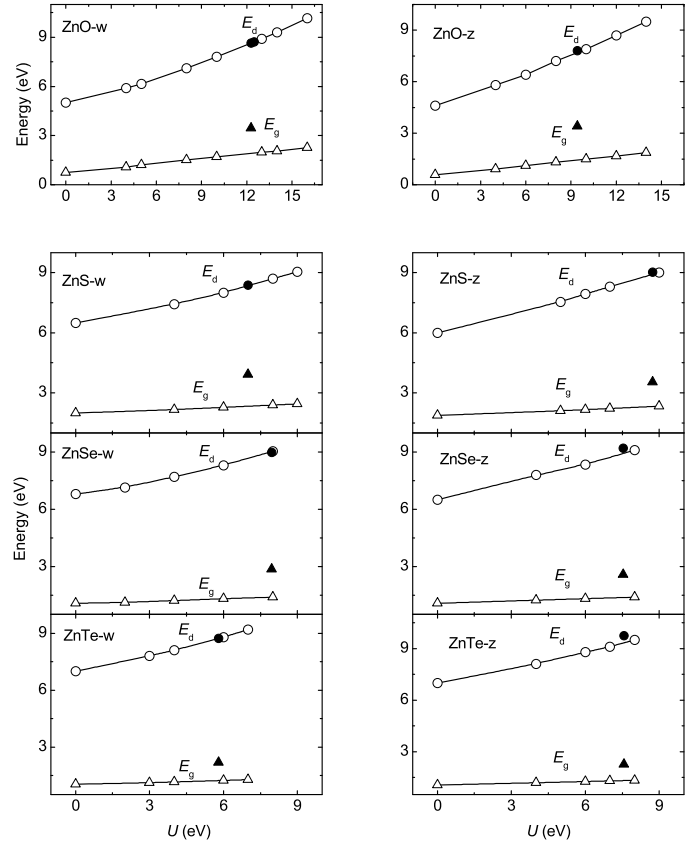}
\setlength{\abovecaptionskip}{0.0cm}
\caption{Band gap ($E_g$) and mean energy level of the
Zn-3$d$~states ($E_d$) relative to the VB maximum for Zn$X$ with
z- and w-type structure as a function of the parameter $U$. Open
symbols correspond to calculated data, and filled symbols are
experimental data from Refs.~\onlinecite{RSS94,HM87,LPMKS74,VL71}.
Due to the lack of experimental data for ZnO-z the experimental
values for ZnO-w from Ref.~\onlinecite{HM87} are used in the top
right panel.} \label{fig:u}
\end{figure}

\noindent
\begin{figure}[]
\centering
\includegraphics[angle=0, width=1.00\textwidth, height=0.90\textwidth]{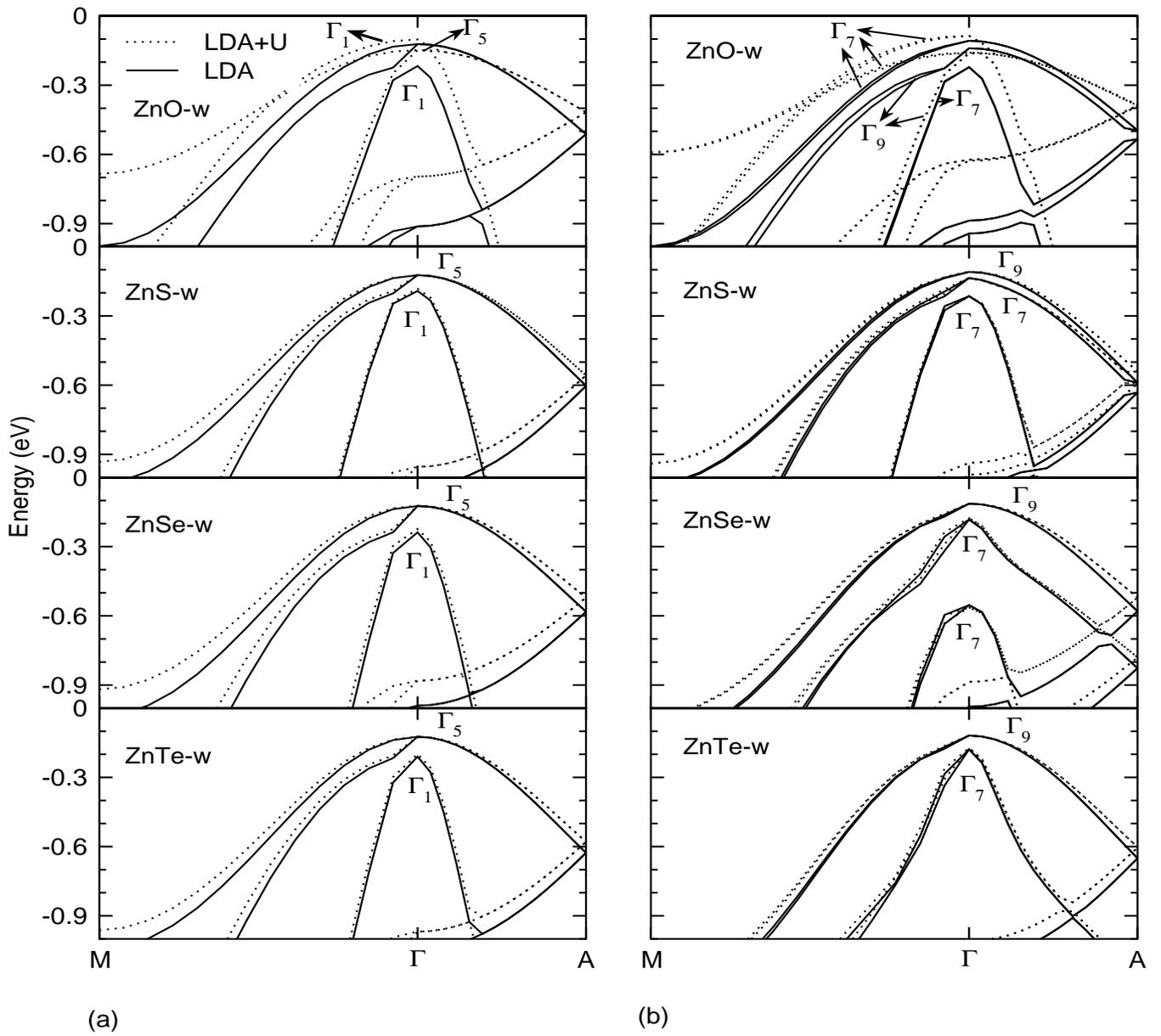}
\caption{Band structure of Zn$X$-w near the VB maximum
calculated by LDA (solid line) and LDA+$U$ (dotted line)
approaches: (a) neglecting and (b) including the SO coupling. }
\label{wFS}
\end{figure}

\begin{figure}[h]
\centering
\includegraphics[angle=0, width=1.00\textwidth, height=0.90\textwidth]{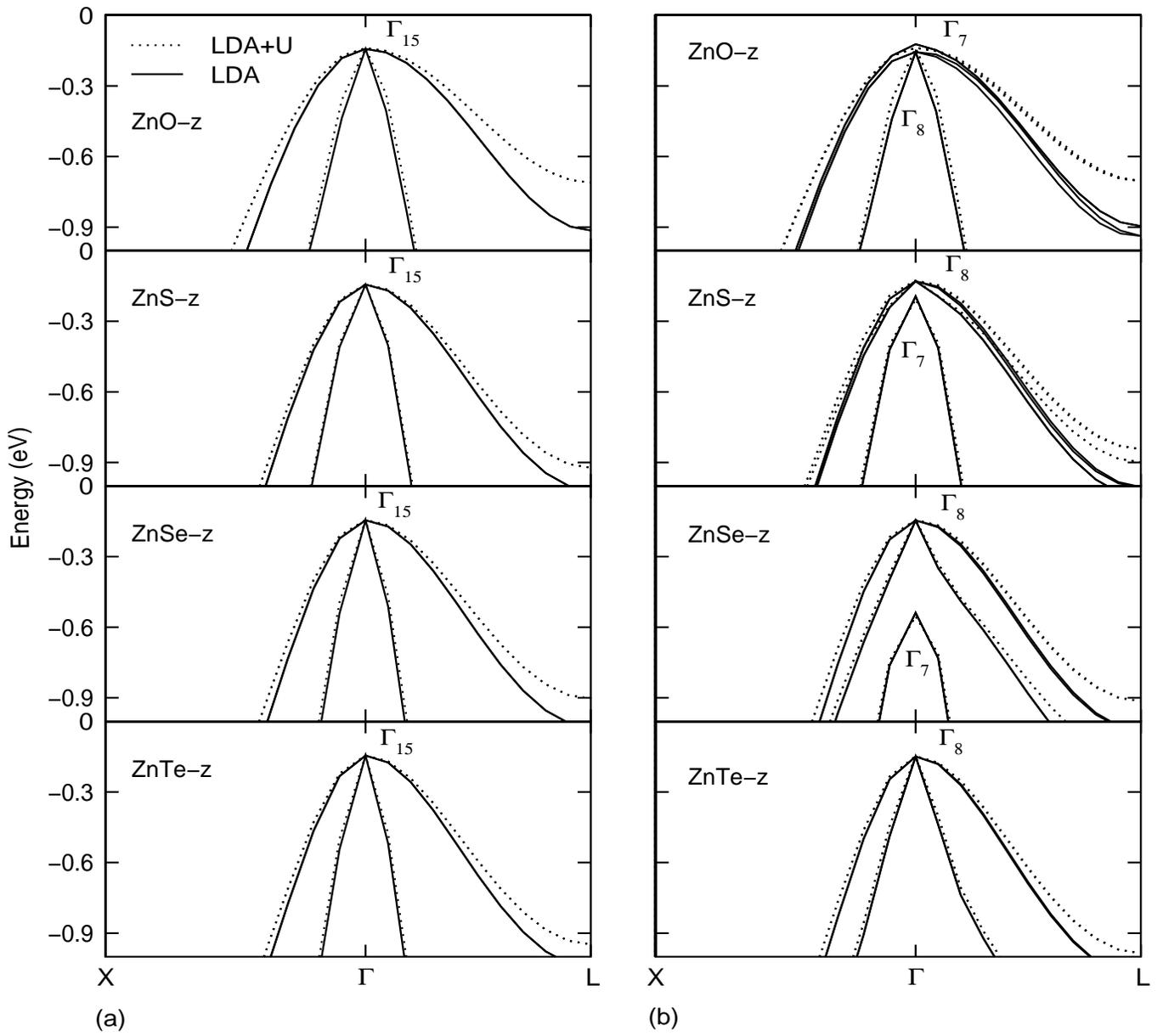}
\caption{Band structure for Zn$X$-z phases calculated by LDA
(solid line) and LDA+$U$ (dotted line) approaches: (a) neglecting
and (b) including the SO coupling. } \label{zFS}
\end{figure}

\begin{figure}[h]
\centering
\includegraphics[angle=0, width=1.00\textwidth, height=0.90\textwidth]{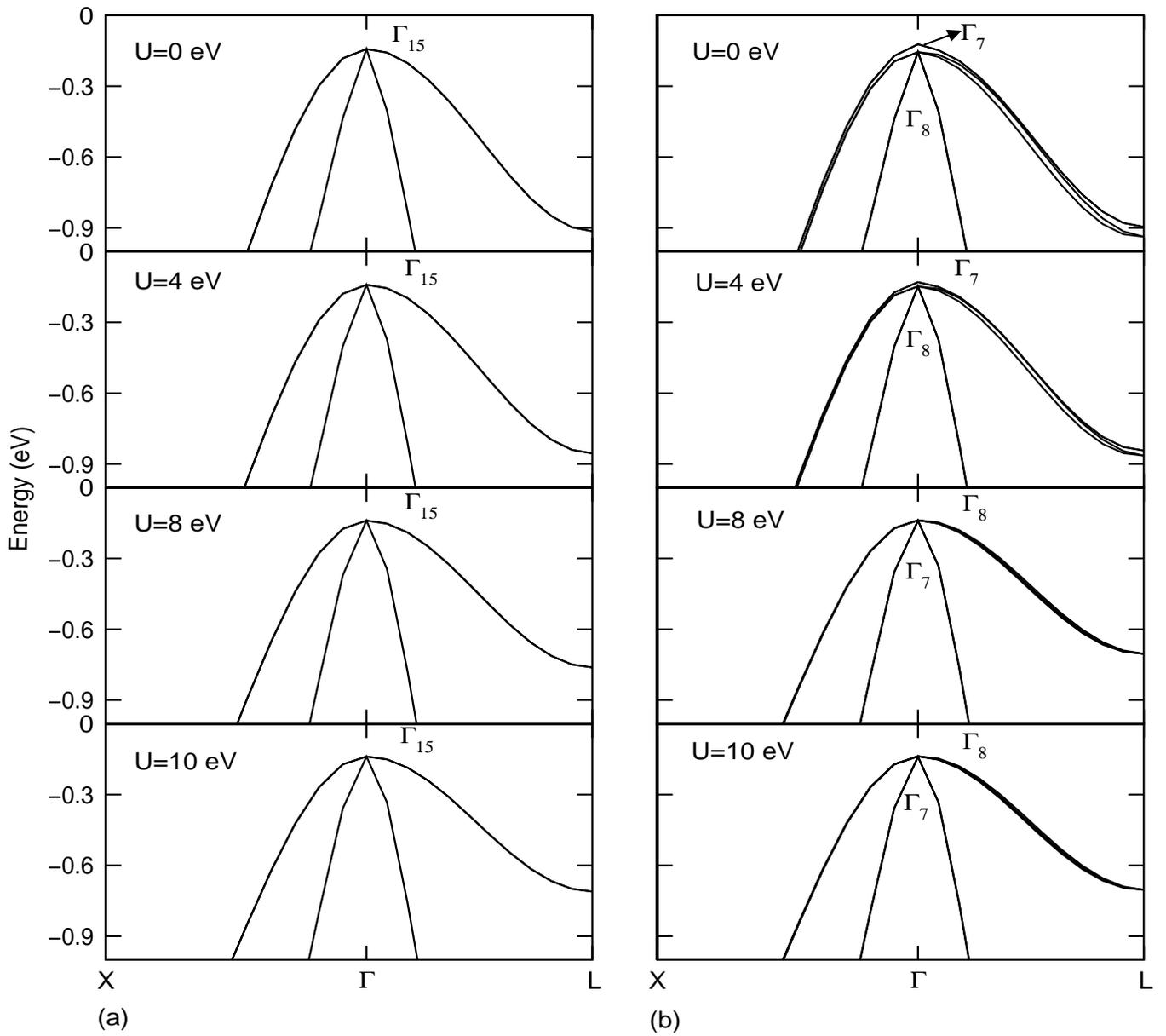}
\caption{Evolution of the band structure of ZnO-z near the VB
maximum with increasing the $U$: (a) neglecting and (b) including
the SO coupling. } \label{zZnO+FS}
\end{figure}

\begin{figure}[h]
\centering
\includegraphics[angle=0, width=1.00\textwidth, height=0.7\textwidth]{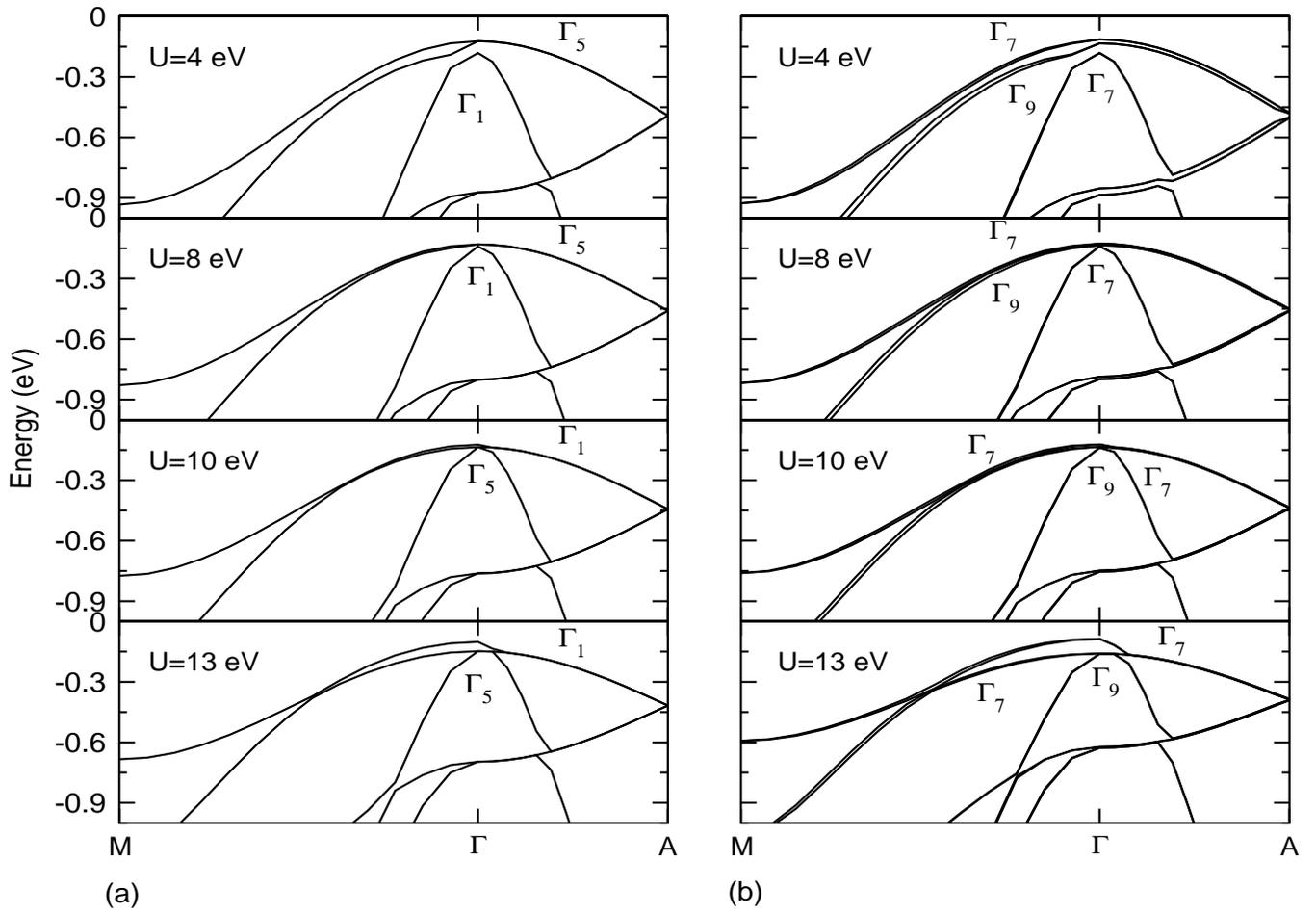}
\caption{Dependence of band structure of ZnO-w near the VB
maximum on the parameter $U$: (a) neglecting and (b) including the
SO coupling. } \label{wZnO+FS}
\end{figure}

\begin{figure}[h]
\centering \vspace{-3.0cm}
\includegraphics[scale=1.0]{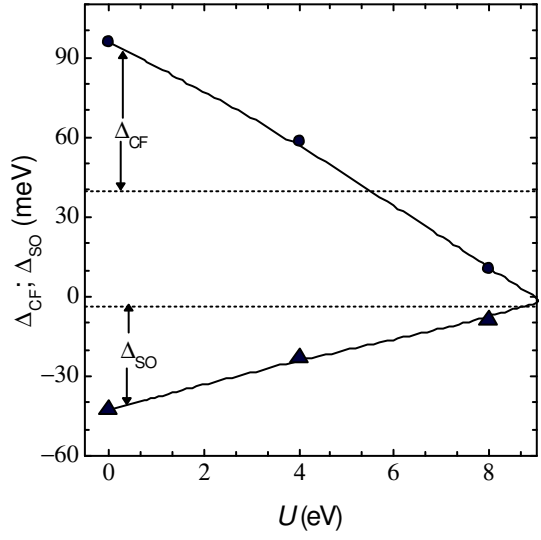}
\caption{Crystal-field and spin-orbit energy splitting as a
function of $U$ for ZnO-w. Solid and dotted lines represent
calculated and experimental data, respectively.} \label{ZnO+CF+SO}
\end{figure}

\begin{figure}[h]
\centering 
\includegraphics[scale=1.0]{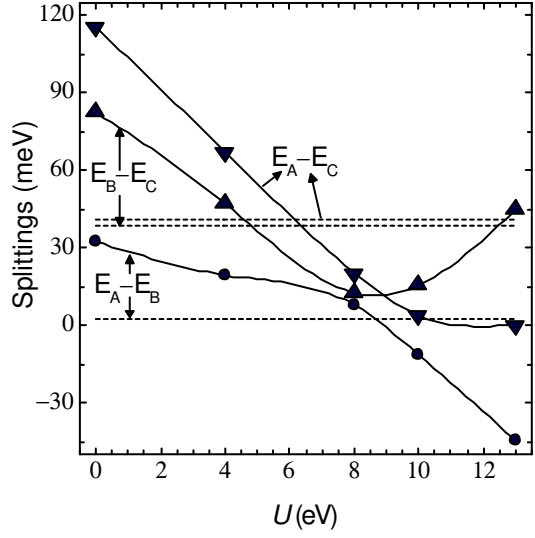} \setlength{\abovecaptionskip}{-11.0cm}
\caption{Splitting of the states at the top of VB vs. $U$ for
ZnO-w. Solid and dotted lines represent calculated and
experimental data, respectively.} \label{ZnO+ABC}
\end{figure}

\begin{figure}[h]
\centering
\vspace{-3.0cm}
\includegraphics[scale=1.0]{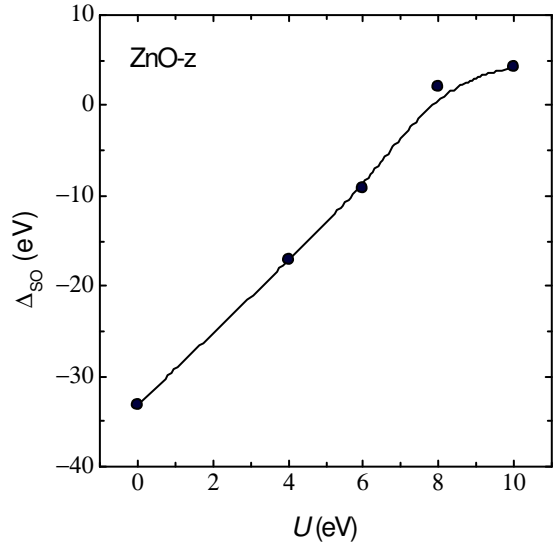}
\setlength{\abovecaptionskip}{-11.0cm} \caption{Spin-orbit
splitting energy for the zinc-blende ZnO.} \label{zZnO+SO}
\end{figure}
}

\end{document}